\documentclass[useAMS,usenatbib]{mn2e}

\usepackage{natbib}
\usepackage{graphicx}
\usepackage{multicol}
\usepackage{longtable}
\usepackage{amssymb}
\usepackage[width=1.0\textwidth]{caption}
\usepackage[usenames,dvipsnames]{color}
\RequirePackage{subfigure} %
\definecolor{Red}{rgb}{1.0,0,0}

\newcommand{\etal}    {{\it et al}}                          
\newcommand{\otp}     {O$^{2+}$}
\newcommand{\SLP} [3]{$^{#1}$#2$^{\rm{#3}}$}            
\newcommand{\SLPJ}[4]{$^{#1}$#2$^{\rm{#3}}_{_{#4}}$}       
\newcommand{\nlo} [3]{#1#2$^{#3}$}                 
\newcommand{\po}  [2]{\overline{#1}\rm{#2}}        
\newcommand{\poi} [2]{\overline{#1}{#2}}
\newcommand{\AS}    {Autostructure\index{Autostructure}}

\newcommand{\II}      {~{\sc ii}}

\newcommand{\Tf}      {T_{\rm f}}

\title[{Collision Strengths for [O~{\sc iii}] Optical and Infrared Lines}]{Collision Strengths for [O~{\sc iii}] Optical and Infrared Lines}
\author[P.J. Storey, Taha Sochi \& N.R. Badnell]{P.J. Storey$^{1}$\thanks{E-mail:pjs@star.ucl.ac.uk (PJS).}, Taha Sochi$^{1}$\footnotemark[1]\thanks{E-mail: t.sochi@ucl.ac.uk (TMS). Corresponding author.} and N.R. Badnell$^2$\footnotemark[1]\thanks{E-mail: badnell@phys.strath.ac.uk (NRB).} \\
$^{1}$University College London, Department of Physics and
Astronomy, Gower Street, London, WC1E 6BT \\
$^2$ Department of Physics, University of Strathclyde, Glasgow, G4 0NG, UK}

\begin{document}

\date{Accepted XXX. Received XXX; in original form XXX}


\maketitle

\label{firstpage}

\begin{abstract}
  We present electron collision strengths and their thermally averaged
  values for the forbidden lines of the astronomically abundant
  doubly-ionized oxygen ion, \otp, in an intermediate coupling scheme
  using the Breit-Pauli relativistic terms as implemented in an
  R-matrix atomic scattering code. We use several atomic targets for
  the R-matrix scattering calculations including one with 72 atomic
  terms. We also compare with new results obtained using the intermediate
  coupling frame transformation method. We find spectroscopically significant
  differences against a recent Breit-Pauli calculation for the
  excitation of the [O~{\sc iii}] $\lambda$4363 transition but confirm
  the results of earlier calculations.
\end{abstract}


\section{Introduction} \label{Introduction}

The forbidden lines of \otp\ are among the most important features in the spectra of photoionized
plasmas which include, {\it inter alia}, H\II\ regions and planetary nebulae. The exceptional
brightness of the strongest [O~{\sc iii}] lines means that they can be used to determine the oxygen
abundances and physical conditions in the Milky Way and other galaxies out to cosmological
distances that reach redshifts of more than $z=3$ \citep{Maiolino2008}.

It has been recently suggested \citep{NichollsDS2012} that the elemental abundance and electron
temperature anomalies seen in the analysis of the planetary nebula spectra, where considerable
differences have been observed between the results obtained from the collisionally-excited lines
(CEL) and those obtained from the optical recombination lines (ORL), might be resolved by using non
Maxwell-Boltzmann (MB) distributions for the energies of the free electrons. The $\kappa$
distribution, which is widely used in the analysis of solar data, was proposed as a replacement for
the MB distribution to resolve this issue. If the electron distributions are generally
non-Maxwellian in nebulae it would affect the analysis of [O~{\sc iii}] lines significantly and
reliable collision strength data are needed to compute the effective collision strengths for
collisional excitation and de-excitation.

The proposal that the electron energy distribution in planetary nebulae is not Maxwellian dates
back to the 1940s at least where \citet{Hagihara1944} proposed that the velocity distribution of
free electrons in gaseous assemblies, such as those found in planetary nebulae, deviates
significantly from the Maxwellian. \citet{BohmA1947} argued against Hagihara and concluded that any
deviation from the Maxwellian equilibrium distribution is very small. The essence of Bohm and
Aller's argument is that for typical planetary nebulae conditions of electron temperature of about
10000~K and electron number density of about $10^4$~cm$^{-3}$, the thermalization process of
elastic collisions between an electron and other electrons and ions is by far the most frequent
event and typically occurs once every second, while other processes that shift the system from its
thermodynamic equilibrium, like inelastic scattering with other ions that leads to metastable
excitation or recapture, occur at much larger time scales estimated to be months or even years.
Bohm and Aller also indicated the significance of any possible deviation from a Maxwellian
distribution on derived elemental abundances.

Although there have been many studies related to collision strengths of O$^{2+}$, as we will
discuss in the coming paragraphs, some of the previous data have limitations. For example, some of
these data are produced in an $LS$ coupling scheme while others are based on approaches that do not
adequately treat resonance phenomena.

Before the advent of close-coupling codes there were several calculations of collision strengths
for excitation of the O~{\sc iii} forbidden lines that did not incorporate resonance effects
\citep{CzyzakKMSSS1968,Seaton1975,BhatiaDF1979}.

The first close-coupled collision strengths were obtained by \citet{BalujaBK1980} for some of the
semi-forbidden intercombination transitions of O~{\sc iii}\ using the R-matrix method
\citep{BerringtonBCCRT1974,BerringtonBBSSTY1987,HummerBEPST1993,BerringtonEN1995}. They included
all channels with configurations \nlo1s2 \nlo2s2 \nlo2p2, \nlo1s2 2s \nlo2p3 and \nlo1s2 \nlo2p4 in
the expansion of the wavefunction. They also used three pseudo-orbitals ($\po3s$, $\po3p$ and
$\po3d$) and allowed for configuration interaction in the included states with the addition of
correlation terms in the total wavefunction.

\citet{HoH1983} also used the close coupling approximation with configuration interaction in the
target wavefunction to compute the collision strengths of some of O~{\sc iii}\ transitions in $LS$
coupling. They employed a mix of spectroscopic and correlation Hartree-Fock orbitals to describe
their target.

Relatively extensive work was done by
\citet{Aggarwal1983,Aggarwal1985} who computed collision strengths of
O~{\sc iii} transitions between the fine structure levels using
configuration interaction target wavefunctions. He transformed $LS$
coupling reactance matrices obtained from R-matrix calculations to
pair coupling with the program JAJOM \citep{Saraph1978}. The results
were obtained with a fine energy mesh up to 5.16~Rydberg where a
complex resonance structure was observed on the entire mesh.

\citet{Aggarwal1993} used an elaborate configuration interaction
target described by \citet{AggarwalH1991} and the R-matrix method in
$LS$ coupling to compute effective collision strengths for some
inelastic transitions of O~{\sc iii}\ between 26 $LS$-coupled states
of six configurations over a wide range of electron temperature
(2500-200000~K). They employed the standard and no-exchange R-matrix
codes on a fine energy mesh that reveals the resonance
structure. This work was extended by \citet{AggarwalK1999},
who they computed the collision strengths for the transitions
between the fine structure levels using the R-matrix method including
all partial waves with $L \le 40$ to ensure convergence.
\citet{AggarwalK1999} transformed the $LS$ reactance matrices obtained
by \citet{Aggarwal1993} into pair coupling using JAJOM
\citep{Saraph1978} where necessary. They only tabulated fine-structure
collision strengths for some transitions, pointing out that in pair
coupling, if one of the terms in a transition has spin zero and hence
$J=L$, e.g. $^3$P -- $^1$D, the fine-structure collision strengths are
proportional to the statistical weight of the non-zero spin states, in
this example the $^3$P$_J$ levels.

\citet{LennonB1994} did extensive work on O$^{2+}$ collision strengths for the transitions
between the fine structure levels, as part of a wider investigation on the carbon isoelectronic
ions, using the R-matrix method, where the CIV3 configuration interaction code \citep{Hibbert1975}
was used to generate the target wavefunctions. The target included 12 states belonging to 3
configurations (\nlo1s2 \nlo2s2 \nlo2p2, \nlo1s2 2s \nlo2p3 and \nlo1s2 \nlo2p4). They also
transformed to pair coupling in the same way as \citet{AggarwalK1999} described above. They
presented a sample of Maxwellian based effective collision strengths in the temperature range
$10^{3}-10^{5}$~K.

Recently, \citet{PalayNPE2012} made the first calculation of collision strengths for the O~{\sc
iii} forbidden transitions using a relativistic Breit-Pauli (BP) R-matrix method with resolved
resonance structures. They used 22 configurations (3 spectroscopic and 19 correlation) to describe
the target. Like most of the previous studies, they have also presented samples of the Maxwellian
averaged effective collision strengths which were also computed at temperatures down to 100~K.

The most recent R-matrix calculations \citep{LennonB1994,AggarwalK1999,PalayNPE2012} generally
agree to within 10\% for the thermally averaged collision strengths for the forbidden transitions
among the five lowest levels. An exception to this generally close agreement is for the transitions
from the lowest three $^3$P$_J$ levels to the $^1$S$_0$ state. The recent results of
\citet{PalayNPE2012} differ significantly from those of earlier workers. The excitation mechanism
of the $^1$S$_0$ level is important because the $^1$S$_0 \rightarrow ^1$D$_2$ $\lambda 4363$ line
is widely used to infer the electron temperature in H~{\sc ii} regions and planetary nebulae. If a
$\kappa$ distribution of electron energies is assumed, the number of free electrons capable of
exciting the $^1$S$_0$ state would be increased relative to a MB distribution which would affect
the derived \otp\ abundance.

The aim of the present paper is twofold. Firstly we make a Breit-Pauli R-matrix calculation of the
\otp\ collision strengths with an independently derived target configuration basis to compare with
previous work, especially the only other Breit-Pauli results from \citet{PalayNPE2012}. Secondly we
attempt to place realistic error estimates on our results by examining the effect of several
factors on our results. We discuss the convergence of our calculation as the number of target
states is increased. Our largest target includes significant contributions to the dipole
polarizability of the three energetically lowest terms. We also consider the effect of Gailitis
averaging of the collision strengths close to the excitation thresholds, especially for excitation
of the $^3$P$_1$ level between the $^3$P$_1$ and $^3$P$_2$ thresholds. We additionally compare the
results of the Breit-Pauli calculation with those obtained using the Intermediate Coupling Frame
Transformation (ICFT) R-matrix method \citep{griffinetal98}. This method is based on transforming
the non-physical $LS$-coupled reactance matrices, to compute collision strengths in intermediate
coupling.

The calculation described in the following sections is constructed to provide accurate results for
the excitation of the optical and infrared forbidden transitions among the five lowest levels of
\otp\ at temperatures typical of photoionized plasmas. We compute collision strengths up to
$\approx 1.3$~Rydberg free electron energy relative to the ground level and Mawell-Boltzmann
averaged collision strengths from 100~K to 25000~K.

The main tools used in this investigation are the \AS\ code\footnote{{See Badnell: Autostructure
write-up on WWW. URL: amdpp.phys.strath.ac.uk/autos/ver/WRITEUP.}}
\citep{EissnerJN1974,NussbaumerS1978,AS2011} to define and elaborate the atomic target and the
UCL-Belfast-Strathclyde R-matrix code\footnote{{See Badnell: R-matrix write-up on WWW. URL:
amdpp.phys.strath.ac.uk/UK\_RmaX/codes/serial/WRITEUP.}} \citep{BerringtonEN1995} to do the actual
scattering calculations. We compare our results with earlier calculations and also assess the
reliability of our results.

\section{Computation}

In the following we outline the computational methods used in this work.

\subsection{The \otp\ target}\label{target}

We used the \AS\ code \citep{AS2011} to generate the target radial functions required as an input
to the first stage of the R-matrix code. The radial data were generated using thirty-nine
configurations containing seven orbitals; three physical (1s, 2s and 2p) and four correlation
orbitals ($\po3s$, $\po3p$, $\po3d$, $\po4f$). These configurations are given in Table
\ref{ConfigTable}. An iterative optimization variational protocol was used to obtain the orbital
scaling parameters, $\lambda_{nl}$, which are given in Table~\ref{lambdaTable}. The correlation
orbitals are calculated in a Coulomb potential with central charge 8$|\lambda_{nl}|$.

\begin{table} 
\caption{The configuration basis used to define the scattering target. The 1s$^2$ core is
to be understood in all configurations. The bar signifies a correlation orbital.} \label{ConfigTable}
\centering
\begin{tabular}{|l|}
\hline
 2s$^2$~2p$^2$, 2s~2p$^3$, 2p$^4$ \\
 2s$^2$~2p~$\poi3l$; $l=0,1,2$ \\
 2s~2p$^2$~$\poi3l$; $l=0,1,2$ \\
 2p$^3$~$\poi3l$; $l=0,1,2$ \\
 2s$^2$~$\poi3l$~$\poi3l^{'}$, $l,l^{'}$=0,1,2 \\
 2s~2p~$\poi3l$~$\poi3l^{'}$, $l,l^{'}$=0,1,2 \\
 2p$^2$~$\poi3l$~$\poi3l^{'}$, $l,l^{'}$=0,1,2 \\
 2s$^2$~2p~$\po4f$, 2s~2p$^2$~$\po4f$, 2p$^3$~$\po4f$ \\
 2s$^2$~$\po3d$~$\po4f$, 2s~2p~$\po3d$~$\po4f$, 2p$^2$~$\po3d$~$\po4f$ \\
 2s$^2$~$\po4f^2$, 2s~2p~$\po4f^2$, 2p$^2$~$\po4f^2$ \\
\hline
\end{tabular}
\end{table}

\begin{table} 
  \caption{Orbital scaling parameters, $\lambda_{nl}$, for
    \AS\ input. The rows stand for the principal quantum number $n$,
    while the columns stand for the orbital angular momentum quantum
    number $l$.} \label{lambdaTable} 
\begin{tabular}{|l|c|c|c|c|c|}
\hline
           &    {\bf s} &    {\bf p} &    {\bf d} &    {\bf f} \\
\hline
   1 &    1.44889 &            &            &            \\
\hline
   2 &    1.22418 &    1.18282 &            &            \\
\hline
   $\overline3$ &   -0.80508 &   -0.61905 &   -1.04731 &            \\
\hline
   $\overline4$ &            &            &            &    -1.87410 \\
\hline
\end{tabular}
\end{table}

\begin{table*} 
\caption{Target terms and energies, $E$, calculated by \AS\ using the configuration basis listed in
Table~\ref{ConfigTable}. The 1s$^2$ core is suppressed from all configurations. All these terms are
included in the 72-term target, while for the smaller targets (10- and 20-term) only the first 10
and 20 terms respectively are included.\label{TermTable}} \vspace{-0.2cm}
\begin{center}
{\small
\begin{tabular}{|cccc|cccc|}
\hline
{\bf Index} & {\bf Configuration} & {\bf Term} & {\bf $E$ (cm$^{-1}$)} & {\bf Index} & {\bf Configuration} & {\bf Term} & {\bf $E$ (cm$^{-1}$)} \\
\hline
         1 & \nlo2s2 \nlo2p2 &     \SLP3Pe &        0.0 &         37 & 2s \nlo2p2 $\po3p$ &     \SLP1So &   483995 \\

         2 & \nlo2s2 \nlo2p2 &     \SLP1De &    21257 &         38 & 2s \nlo2p2 $\po3p$ &     \SLP3Po &   485363 \\

         3 & \nlo2s2 \nlo2p2 &     \SLP1Se &    45630 &         39 & 2s \nlo2p2 $\po3s$ &     \SLP3Se &   486363 \\

         4 & 2s \nlo2p3 &     \SLP5So &    58948 &         40 & 2s \nlo2p2 $\po3p$ &     \SLP1Po &   486594 \\

         5 & 2s \nlo2p3 &     \SLP3Do &   121133 &         41 & 2s \nlo2p2 $\po3p$ &     \SLP3Do &   491627 \\

         6 & 2s \nlo2p3 &     \SLP3Po &   144640 &         42 & 2s \nlo2p2 $\po3s$ &     \SLP3Pe &   496357 \\

         7 & 2s \nlo2p3 &     \SLP1Do &   190314 &         43 & 2s \nlo2p2 $\po3p$ &     \SLP3Po &   498566 \\

         8 & 2s \nlo2p3 &     \SLP3So &   199693 &         44 & 2s \nlo2p2 $\po3p$ &     \SLP3So &   503468 \\

         9 & 2s \nlo2p3 &     \SLP1Po &   214885 &         45 & 2s \nlo2p2 $\po3s$ &     \SLP1Pe &   506149 \\

        10 &    \nlo2p4 &     \SLP3Pe &   287613 &         46 & 2s \nlo2p2 $\po3s$ &     \SLP1Se &   508564 \\

        11 & \nlo2s2 2p $\po3p$ &     \SLP1Pe &   301182 &         47 & 2s \nlo2p2 $\po3p$ &     \SLP1Do &   509665 \\

        12 &    \nlo2p4 &     \SLP1De &   302782 &         48 & 2s \nlo2p2 $\po3p$ &     \SLP1Po &   526999 \\

        13 & \nlo2s2 2p $\po3p$ &     \SLP3De &   306265 &         49 & \nlo2s2 2p $\po3d$ &     \SLP3Fo &   539361 \\

        14 & \nlo2s2 2p $\po3s$ &     \SLP3Po &   309248 &         50 & \nlo2p3 $\po3p$ &     \SLP5Pe &   542552 \\

        15 & \nlo2s2 2p $\po3p$ &     \SLP3Se &   310788 &         51 & \nlo2s2 2p $\po3d$ &     \SLP1Do &   550865 \\

        16 & \nlo2s2 2p $\po3p$ &     \SLP3Pe &   315730 &         52 & \nlo2p3 $\po3s$ &     \SLP5So &   551832 \\

        17 & \nlo2s2 2p $\po3s$ &     \SLP1Po &   323156 &         53 & \nlo2p3 $\po3p$ &     \SLP3Pe &   553504 \\

        18 & \nlo2s2 2p $\po3p$ &     \SLP1De &   328464 &         54 & \nlo2p3 $\po3p$ &     \SLP3De &   567272 \\

        19 & \nlo2s2 2p $\po3p$ &     \SLP1Se &   345567 &         55 & \nlo2p3 $\po3p$ &     \SLP1Pe &   567946 \\

        20 &    \nlo2p4 &     \SLP1Se &   351203 &         56 & \nlo2p3 $\po3p$ &     \SLP3Fe &   568179 \\

        21 & 2s \nlo2p2 $\po3p$ &     \SLP3So &   372370 &         57 & \nlo2p3 $\po3p$ &     \SLP1Fe &   570684 \\

        22 & 2s \nlo2p2 $\po3p$ &     \SLP5Do &   376169 &         58 & \nlo2s2 2p $\po3d$ &     \SLP3Po &   572129 \\

        23 & 2s \nlo2p2 $\po3s$ &     \SLP5Pe &   379161 &         59 & \nlo2s2 2p $\po3d$ &     \SLP3Do &   578854 \\

        24 & 2s \nlo2p2 $\po3p$ &     \SLP5Po &   379976 &         60 & \nlo2p3 $\po3s$ &     \SLP3Do &   584855 \\

        25 & 2s \nlo2p2 $\po3p$ &     \SLP3Do &   392038 &         61 & \nlo2p3 $\po3s$ &     \SLP3So &   589631 \\

        26 & 2s \nlo2p2 $\po3p$ &     \SLP5So &   395094 &         62 & \nlo2p3 $\po3p$ &     \SLP3Pe &   600736 \\

        27 & 2s \nlo2p2 $\po3p$ &     \SLP3Po &   400796 &         63 & 2s \nlo2p2 $\po3d$ &     \SLP5Fe &   600824 \\

        28 & 2s \nlo2p2 $\po3s$ &     \SLP3Pe &   418063 &         64 & \nlo2p3 $\po3s$ &     \SLP1Do &   602294 \\

        29 & 2s \nlo2p2 $\po3p$ &     \SLP3Fo &   437947 &         65 & \nlo2p3 $\po3p$ &     \SLP1De &   606079 \\

        30 & 2s \nlo2p2 $\po3p$ &     \SLP1Do &   439915 &         66 & \nlo2p3 $\po3p$ &     \SLP3Se &   606299 \\

        31 & 2s \nlo2p2 $\po3s$ &     \SLP3De &   442045 &         67 & \nlo2p3 $\po3p$ &     \SLP3De &   607292 \\

        32 & 2s \nlo2p2 $\po3p$ &     \SLP1Fo &   443431 &         68 & 2s \nlo2p2 $\po3d$ &     \SLP5De &   610073 \\

        33 & 2s \nlo2p2 $\po3p$ &     \SLP3Do &   447068 &         69 & \nlo2p3 $\po3p$ &     \SLP1Pe &   612417 \\

        34 & 2s \nlo2p2 $\po3p$ &     \SLP1Po &   449383 &         70 & \nlo2s2 2p $\po3d$ &     \SLP1Po &   617417 \\

        35 & 2s \nlo2p2 $\po3p$ &     \SLP3Po &   456397 &         71 & \nlo2p3 $\po3p$ &     \SLP3Pe &   618955 \\

        36 & 2s \nlo2p2 $\po3s$ &     \SLP1De &   468076 &         72 & \nlo2s2 2p $\po3d$ &     \SLP1Fo &   620437 \\
\hline
\end{tabular}
}
\end{center}
\end{table*}

In the scattering calculations, targets with differing numbers of target states were used, with the
largest having 72 terms which are listed in Table~\ref{TermTable}. Calculations were also made with
the first 10 and 20 terms from this list as discussed in more detail below.

In Table~\ref{gfTable} we show the statistically weighted oscillator strengths, $gf$, in the length
and velocity formulations for all the strong allowed transitions between the 2s$^2$~2p$^2$ and
2s~2p$^3$ configurations. The agreement is excellent. Good agreement between the length and
velocity results is a necessary but not sufficient condition for ensuring the quality of the target
wave functions. These transitions also make the largest contributions to the dipole
polarizabilities of the three lowest terms.

The 72 terms listed in Table~\ref{TermTable} were chosen to include all those correlation
configurations that contribute significantly to the dipole polarizability of the three lowest
terms. The main contributions come from 2s$^2$~2p~$\po3d$ configuration. The contribution of states
outside the $n=2$ complex to the dipole polarizabilities of the $^3$P, $^1$D and $^1$S terms is
37\%, 37\% and 60\% respectively. In Table~\ref{levelsTable} we list the energies of the 18 levels
of the $n=2$ complex configurations. We show theoretical energies which include one- and two-body
fine-structure interactions ($E{_{\rm{th2}}}$) and those which only include the spin-orbit
interaction ($E{_{\rm{th1}}}$), the latter being the only fine-structure interactions included in
the version of the R-matrix code that we use (see footnote 2). We
return to the effect of omitting two-body fine structure interactions in section~\ref{Results}.

\subsection{The Scattering Calculations}\label{scattering}

We made several calculations with increasing numbers of target states, both with Breit-Pauli and
the Intermediate Coupling Frame Transformation R-matrix methods. The target radial functions were
supplied as a radial grid format rather than Slater type orbital format where the radial file was
generated by \AS. The inner region radius (RA) in the R-matrix formulation was 9.315~au. Twelve
continuum basis functions were used to represent the wavefunctions in the inner region. This choice
was based on convergence tests and with experience from previous work on the C$^+$+e system
\citep{SochiThesis2012,SochiSCIIList2013}. The maximum value of $2J$ for the $(N+1)$-electron
problem was chosen to be 19 although 15 was found to be sufficient for convergence of the collision
strengths for the forbidden transitions of interest here.

As indicated previously, we made three sets of calculations using the configuration basis described in section~\ref{target}
with 10-, 20- and 72-terms using both the BP and ICFT approaches. These three targets comprise 18,
34 and 146 fine-structure levels respectively. The main purpose of using several targets is to have
an estimate of the error in the final results from observing the convergence of the results with
different numbers of target terms. For all three targets, the $(N+1)$-electron wavefunction
contains all possible configurations formed from the 39 configurations of the $N$-electron target
combined with any of the orbitals, spectroscopic and correlation.  There are 102 such
$(N+1)$-electron configurations.

Experimental energies obtained from the National Institute of Standards and Technology
(NIST)\footnote{{See NIST website: www.nist.gov.}} were used in place of theoretical ones to ensure
correct positioning of thresholds for convergence of resonance series. In some cases this required re-ordering the target states.


\begin{table} 
\caption{The weighted $LS$ oscillator strengths in length, $(gf)_{_L}$, and velocity, $(gf)_{_V}$,
forms for the transitions within the $n=2$ complex.} \label{gfTable} \vspace{0.2cm} \centering
\begin{tabular}{cc|c|c}
\hline
           &            & {\bf $(gf)_{_L}$} & {\bf $(gf)_{_V}$} \\
\hline
{\bf \SLP3Pe} & {\bf \SLP3Do} &       1.01 &       0.99 \\

    {\bf } & {\bf \SLP3Po} &       1.28 &       1.31 \\

    {\bf } & {\bf \SLP3So} &       1.70 &       1.61 \\

{\bf \SLP1De} & {\bf \SLP1Do} &       1.57 &       1.61 \\

    {\bf } & {\bf \SLP1Po} &       1.18 &       1.22 \\

{\bf \SLP1Se} & {\bf \SLP1Po} &       0.27 &       0.27 \\
\hline
\end{tabular}
\end{table}

\begin{table} 
\caption{The 18 lowest energy levels of \otp\ and their experimental ($E_{\rm{ex}}$) and
theoretical ($E{_{\rm{th1}}}$ and $E{_{\rm{th2}}}$) energies in wavenumbers (cm$^{-1}$). The
experimental energies are obtained from the NIST database while the theoretical energies were
obtained from \AS\ with the configuration basis listed in Table~\ref{ConfigTable}. The energies
$E{_{\rm{th1}}}$ were obtained with only spin-orbit terms in the target Hamiltonian while
$E{_{\rm{th2}}}$ also include two-body fine-structure interactions within the $n=2$ complex.}
\label{levelsTable} \vspace{0.2cm} \centering
\begin{tabular}{|l|l|l|l|l|}
\hline
{\bf Index} & {\bf Level} & {\bf $E_{\rm{ex}}$} & {\bf $E{_{\rm{th1}}}$} & {\bf $E{_{\rm{th2}}}$} \\
\hline
   {\bf 1} & \nlo2s2 \nlo2p2 \SLPJ3Pe0 &       0.00 &       0 &       0 \\

   {\bf 2} & \nlo2s2 \nlo2p2 \SLPJ3Pe1 &     113.18 &     115 &     113 \\

   {\bf 3} & \nlo2s2 \nlo2p2 \SLPJ3Pe2 &     306.17 &     339 &     308 \\

   {\bf 4} & \nlo2s2 \nlo2p2 \SLPJ1De2 &   20273.27 &   21489 &   21471 \\

   {\bf 5} & \nlo2s2 \nlo2p2 \SLPJ1Se0 &   43185.74 &   45900 &   45882 \\

   {\bf 6} & 2s \nlo2p3 \SLPJ5So2 &   60324.79 &   59600 &   59582 \\

   {\bf 7} & 2s \nlo2p3 \SLPJ3Do3 &  120058.2 &  121800 &  121775 \\

   {\bf 8} & 2s \nlo2p3 \SLPJ3Do2 &  120053.4 &  121804 &  121799 \\

   {\bf 9} & 2s \nlo2p3 \SLPJ3Do1 &  120025.2 &  121812 &  121805 \\

  {\bf 10} & 2s \nlo2p3 \SLPJ3Po2 &  142393.5 &  145316 &  145307 \\

  {\bf 11} & 2s \nlo2p3 \SLPJ3Po1 &  142381.8 &  145321 &  145307 \\

  {\bf 12} & 2s \nlo2p3 \SLPJ3Po0 &  142381.0 &  145331 &  145319 \\

  {\bf 13} & 2s \nlo2p3 \SLPJ1Do2 &  187054.0 &  191025 &  191006 \\

  {\bf 14} & 2s \nlo2p3 \SLPJ3So1 &  197087.7 &  200423 &  200405 \\

  {\bf 15} & 2s \nlo2p3 \SLPJ1Po1 &  210461.8 &  215609 &  215591 \\

  {\bf 16} & \nlo2p4 \SLPJ3Pe2 &  283759.7 &  288653 &  288629 \\

  {\bf 17} & \nlo2p4 \SLPJ3Pe1 &  283977.4 &  288863 &  288854 \\

  {\bf 18} & \nlo2p4 \SLPJ3Pe0 &  284071.9 &  288965 &  288951 \\
\hline
\end{tabular}
\end{table}

\newcommand{\Hs}      {\hspace{-0.8cm}} %
\newcommand{\CIF}     {\centering \includegraphics[width=1.85in]} %
\newcommand{\Vmin}    {\vspace{-0.2cm}} %


\begin{figure}
\centering %
\subfigure[1-2]%
{\begin{minipage}[b]{0.5\textwidth} \centering \includegraphics[height=4cm, width=8.5cm]
{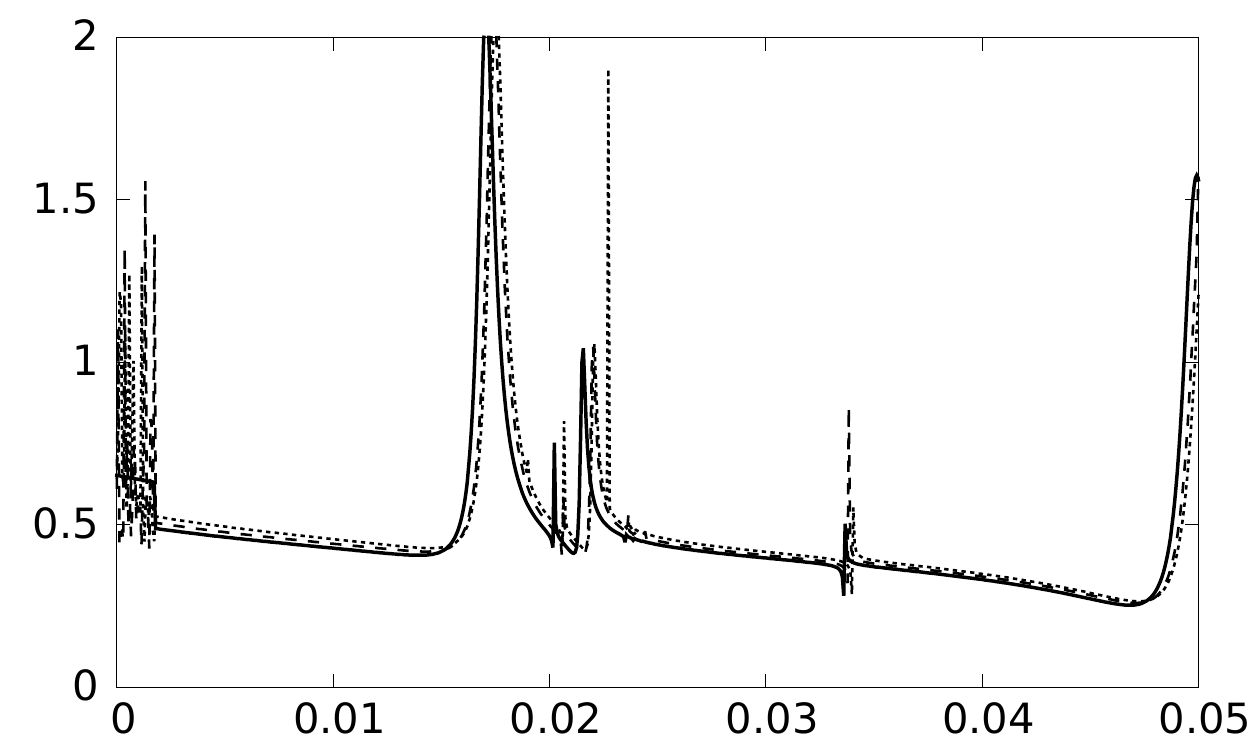}
\end{minipage}} \vspace{-0.3cm}
\centering %
\subfigure[1-3]%
{\begin{minipage}[b]{0.5\textwidth} \centering \includegraphics[height=4cm, width=8.5cm]
{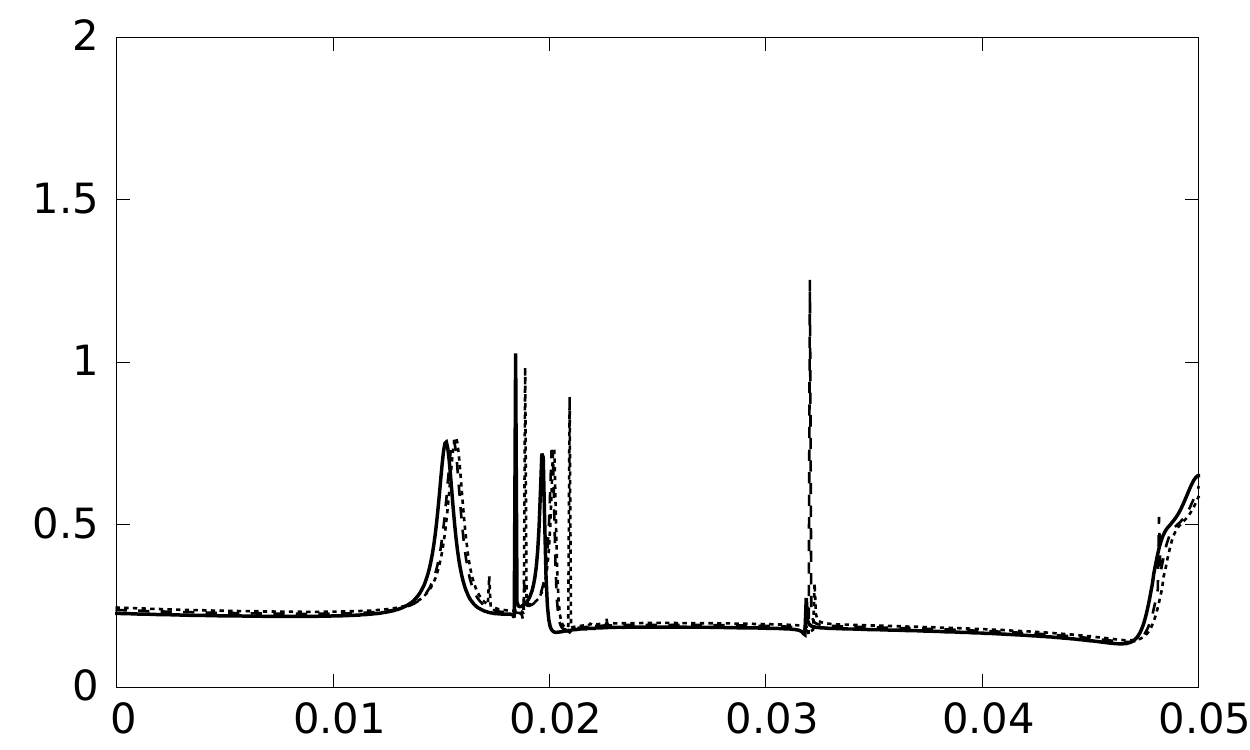}
\end{minipage}} \vspace{-0.3cm}
\centering %
\subfigure[1-4]%
{\begin{minipage}[b]{0.5\textwidth} \centering \includegraphics[height=4cm, width=8.5cm]
{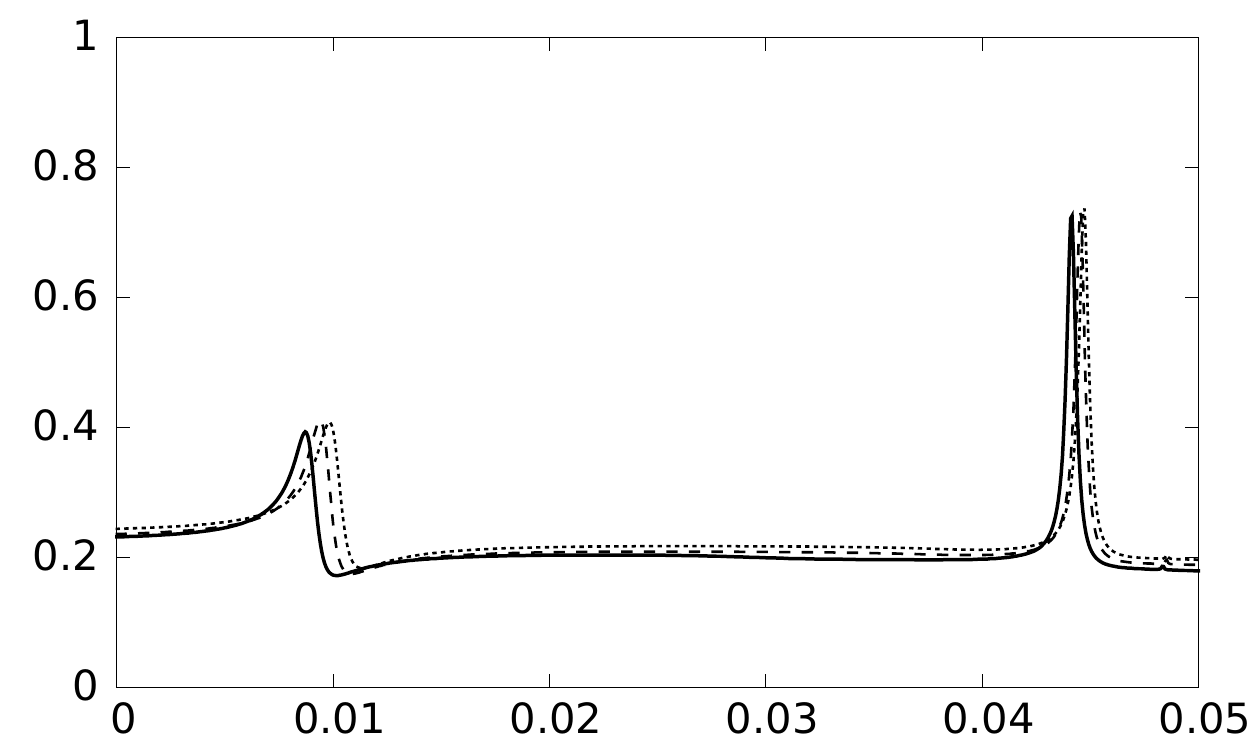}
\end{minipage}}
\caption{Collision strength (vertical axis) versus final electron energy in Rydberg (horizontal
axis) for Breit-Pauli calculations of the 1-2, 1-3 and 1-4 transitions with 10-term (dotted),
20-term (dashed), and 72-term (solid). Refer to Table \ref{levelsTable} for level indexing. \label{CompareOmega}}
\end{figure}

Collision strengths were calculated for electron energies up to 1.28~Rydberg relative to the
2s$^2$~2p$^2$~\SLPJ3Pe0 ground level, hence 0.89~Rydberg relative to the highest state of interest,
2s$^2$~2p$^2$~\SLPJ1Se0. This energy corresponds to $\approx7kT$ when the electron temperature
$T=20000$~K, the approximate upper limit for temperatures in photoionized nebulae. Over this energy
range, collision strengths were calculated at 20000 equally spaced energies, except between the
2s$^2$~2p$^2$~\SLPJ3Pe1 and \SLPJ3Pe2 levels where calculations were performed on a mesh 100 times
finer. Calculations were also made with and without Gailitis averaging of the collision strengths
in the region beneath each threshold where the effective quantum number $\nu>10$.

We calculate the thermodynamically-averaged collision strengths for electron excitation,
$\Upsilon$, from a lower state $i$ to an upper state $j$ from

{\scriptsize
\begin{equation}\label{upsilon}
\Upsilon_{i\rightarrow j}\left(\epsilon_{i},\Tf\right)=\frac{\sqrt{\pi}}{2}e^{\left(\frac{\Delta
E_{ij}}{k\Tf}\right)}\int_{0}^{\infty}\Omega_{ij}\left(\epsilon_{i}\right)\left(\frac{k\Tf}
{\epsilon_{i}}\right)^{1/2}f\left(\epsilon_{i},\Tf\right)d\epsilon_{j}
\end{equation}}
where $\Tf$ is the effective temperature, $k$ is the Boltzmann constant, $\epsilon_{i}$ and
$\epsilon_{j}$ are the free electron energy relative to the states $i$ and $j$ respectively,
$\Delta E_{ij}$ ($=\epsilon_{j}-\epsilon_{i}$) is the energy difference between the two states,
$\Omega_{ij}$ is the collision strength of the transition between the $i$ and $j$ states, and
$f(\epsilon_{i},\Tf)$ is the energy- and temperature-dependent electron distribution.
In what follows we will only consider Maxwell-Boltzmann distributions of electron energy, given by

\begin{equation}
 f_{{\rm MB}}(\epsilon,T)=\frac{2}{\left(kT\right)^{3/2}}\sqrt{\frac{\epsilon}{\pi}}e^{-\frac{\epsilon}{kT}},\label{ElDiEq2}
\end{equation}
although, as discussed above, other distributions such as the $\kappa$ distribution have been
proposed and discussed \citep{Vasyliunas1968, NichollsDS2012, StoreySTemp2013}.

\begin{figure*}
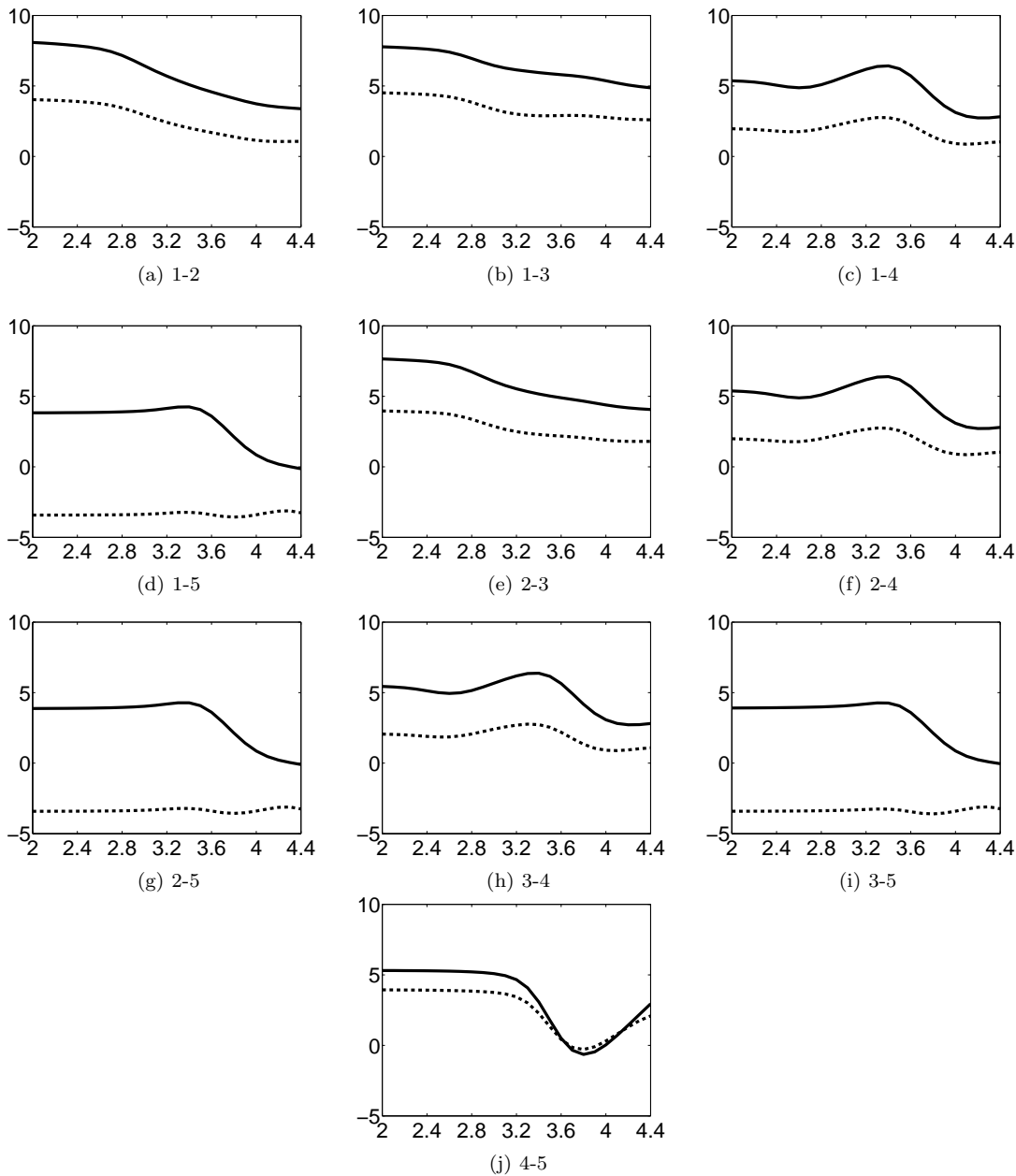

\centering
\subfigure[1-2]%
{
\begin{minipage}[b]{0.32\textwidth}
\CIF {g/BPUpPD1}
\end{minipage}}
\Hs %
\subfigure[1-3]%
{
\begin{minipage}[b]{0.32\textwidth}
\CIF {g/BPUpPD2}
\end{minipage}}
\Hs
\subfigure[1-4]%
{
\begin{minipage}[b]{0.32\textwidth}
\CIF {g/BPUpPD3}
\end{minipage}} \Vmin
%
\centering %
\subfigure[1-5]%
{
\begin{minipage}[b]{0.32\textwidth}
\CIF {g/BPUpPD4}
\end{minipage}}
\Hs
\subfigure[2-3]%
{
\begin{minipage}[b]{0.32\textwidth}
\CIF {g/BPUpPD6}
\end{minipage}}
\Hs
\subfigure[2-4]%
{\begin{minipage}[b]{0.32\textwidth} \CIF {g/BPUpPD7}
\end{minipage}} \Vmin
%
\centering %
\subfigure[2-5]%
{\begin{minipage}[b]{0.32\textwidth} \CIF {g/BPUpPD8}
\end{minipage}} \Vmin
\Hs
\subfigure[3-4]%
{\begin{minipage}[b]{0.32\textwidth} \CIF {g/BPUpPD10}
\end{minipage}}
\Hs
\subfigure[3-5]%
{\begin{minipage}[b]{0.32\textwidth} \CIF {g/BPUpPD11}
\end{minipage}} \Vmin
%
\centering %
\subfigure[4-5]%
{\begin{minipage}[b]{0.32\textwidth} \CIF {g/BPUpPD13}
\end{minipage}}
\caption{Thermally averaged collision strengths from the 10-term target (solid line)
and 20-term target (dotted line), shown as the percentage difference from the 72-term target,
plotted against log~$T$~[K]. The labels of the sub-figures refer to the level indices in Table
\ref{levelsTable}. \label{BPUpPDFig}}
\end{figure*}

\section{Results and Discussion}\label{Results}
\subsection{Results}
A sample of our 10-, 20- and 72-term BP collision strengths is shown in Figure~\ref{CompareOmega}.
The agreement between the three calculations is excellent, with the most obvious difference being
that some resonances move to lower energies as the target size is increased, as might be expected.
Figure~\ref{BPUpPDFig} shows the results for the thermally averaged collision strength, $\Upsilon$,
as a function of temperature for the 10- and 20-term calculations relative to the 72-term
calculation as a percentage difference. The differences are less than 9\% at any temperature for
the 10-term calculation and less than 5\% for the 20-term case.

The energy region between the 2s$^2$~2p$^2$~\SLPJ3Pe1 and \SLPJ3Pe2 states contains a Rydberg
series of resonances converging on the \SLPJ3Pe2 level with an effective quantum number at the
\SLPJ3Pe1 threshold of $47.7$. The energy difference between the \SLPJ3Pe1 and \SLPJ3Pe2 levels of
193~cm$^{-1}$ corresponds to a temperature of 278~K, so this energy region is significant for
computing $\Upsilon$ at temperatures down to 100~K. We calculate the collision strengths with an
energy interval of $6.4\times 10^{-7}$~Rydberg in this interval and compare with the result of
using Gailitis averaging in this region. The difference is less than 1\% at any temperature and we
conclude that Gailitis averaging is adequate to obtain accurate values of $\Upsilon$ down to 100~K.

\begin{figure}
\centering{}
\includegraphics[height=6cm, width=8.5cm]{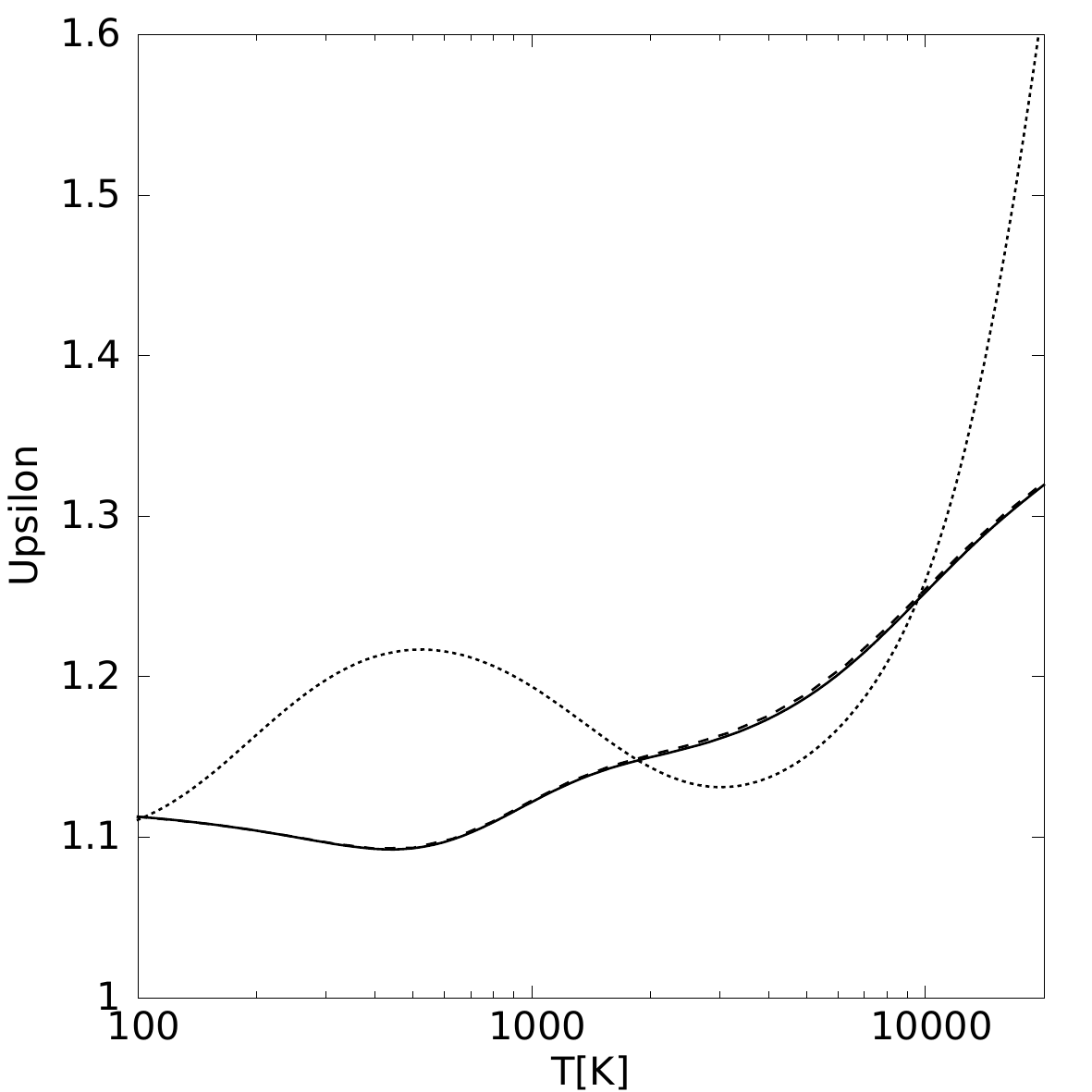}
\caption{Effective collision strength (Upsilon) versus temperature
  from the 72-term target: Breit-Pauli (solid line), ICFT before
  modification (dotted line) and ICFT after modification (dashed
  line), for the transition between levels 2 and 3 of Table
  \ref{levelsTable}. On this scale the solid and dashed lines are
  almost indistinguishable.}  \label{compbpicft}
\end{figure}

The results of the ICFT calculations showed unexpectedly large differences from the BP results in
some energy domains. This is illustrated in Figure~\ref{compbpicft} where we compare the thermally
averaged collision strengths for the 72-term ICFT calculations with the 72-term BP results for the
\SLPJ3Pe1 -- \SLPJ3Pe2 transition.
Due to the difference in scaling with effective charge ($z_{\rm eff}$) of term energy separations
($\propto z_{\rm eff}$) and resonance energies ($\propto z_{\rm eff}^2$), resonance effective
quantum numbers can become small for lowly ionized systems. Such deeply-closed channels can be
problematic for the multi-channel quantum defect theory (MQDT) used by the ICFT method due to
computational finite numerical precision of highly divergent wavefunctions.  \citet{GB2000} found
that classically forbidden channels (e.g. $n<l$) could be handled expediently by simply omitting
them from the MQDT representation. For low-energy scattering in O$^{2+}$ we encountered a similar
problem in a new guise for $n \lesssim 2$. The closed-channel partition of the MQDT representation
should give no contribution since all bound orbitals (spectroscopic and pseudo) are projected out
of the continuum basis. All such closed channel contributions (e.g. correlation resonances) arise
instead in the open-open part of the scattering matrix. For $l>1$ the original \citet{GB2000}
expediency already omits such closed channels ($n<l$). For $l=0,1$ we found it necessary to
explicitly omit such closed channels from the closed partition as well. We show the effect of this
modification as the dashed line in Figure~\ref{compbpicft}. The agreement with the full Breit-Pauli
calculation is now excellent.

Considering the convergence as the number of target states is increased and the good agreement
between the ICFT and Breit-Pauli results, we adopt the results of the 72-term Breit-Pauli
calculation as our final results and, based on the convergence behavior and the effect of Gailitis
averaging, estimate an uncertainty of no more than 5\% in the final thermally averaged collision
strengths. In Table~\ref{UpsilonTable} we tabulate thermally averaged collision strengths
$\Upsilon$, for the 72-term target in the temperature range log$_{10}T$ = 2.0(0.1)4.4.


\begin{table*}
\caption{Thermally averaged collision strengths from the 72-term Breit-Pauli calculation as a
function of temperature. The transitions are  indexed as in Table \ref{levelsTable}.}
\label{UpsilonTable} \centering
\begin{tabular}{|r|cccccccccc|}
\hline
{\bf log~$T$~[K]} &   {\bf 1-2} &   {\bf 1-3} &   {\bf 1-4} &   {\bf 1-5} &   {\bf 2-3} &   {\bf 2-4} &   {\bf 2-5} &   {\bf 3-4} &   {\bf 3-5} &   {\bf 4-5} \\
\hline
 {\bf 2.0} &    0.63500 &    0.22586 &    0.23177 &    0.02989 &    1.11212 &    0.69748 &    0.09003 &    1.17021 &    0.15117 &    0.38267 \\

 {\bf 2.1} &    0.62649 &    0.22561 &    0.23211 &    0.02989 &    1.10993 &    0.69849 &    0.09003 &    1.17188 &    0.15116 &    0.38291 \\

 {\bf 2.2} &    0.61540 &    0.22529 &    0.23259 &    0.02989 &    1.10711 &    0.69994 &    0.09001 &    1.17427 &    0.15112 &    0.38316 \\

 {\bf 2.3} &    0.60216 &    0.22490 &    0.23335 &    0.02988 &    1.10362 &    0.70220 &    0.08998 &    1.17802 &    0.15106 &    0.38346 \\

 {\bf 2.4} &    0.58741 &    0.22447 &    0.23449 &    0.02987 &    1.09952 &    0.70562 &    0.08994 &    1.18370 &    0.15099 &    0.38382 \\

 {\bf 2.5} &    0.57193 &    0.22411 &    0.23600 &    0.02985 &    1.09530 &    0.71013 &    0.08989 &    1.19123 &    0.15089 &    0.38428 \\

 {\bf 2.6} &    0.55668 &    0.22408 &    0.23761 &    0.02983 &    1.09235 &    0.71497 &    0.08982 &    1.19932 &    0.15075 &    0.38486 \\

 {\bf 2.7} &    0.54292 &    0.22475 &    0.23888 &    0.02980 &    1.09280 &    0.71876 &    0.08972 &    1.20571 &    0.15058 &    0.38561 \\

 {\bf 2.8} &    0.53194 &    0.22633 &    0.23934 &    0.02976 &    1.09839 &    0.72014 &    0.08960 &    1.20807 &    0.15036 &    0.38660 \\

 {\bf 2.9} &    0.52440 &    0.22869 &    0.23872 &    0.02970 &    1.10890 &    0.71829 &    0.08942 &    1.20503 &    0.15007 &    0.38792 \\

 {\bf 3.0} &    0.51992 &    0.23126 &    0.23701 &    0.02963 &    1.12183 &    0.71317 &    0.08919 &    1.19654 &    0.14969 &    0.38974 \\

 {\bf 3.1} &    0.51730 &    0.23343 &    0.23441 &    0.02952 &    1.13386 &    0.70539 &    0.08887 &    1.18365 &    0.14917 &    0.39242 \\

 {\bf 3.2} &    0.51544 &    0.23489 &    0.23123 &    0.02937 &    1.14303 &    0.69591 &    0.08842 &    1.16800 &    0.14846 &    0.39684 \\

 {\bf 3.3} &    0.51396 &    0.23583 &    0.22786 &    0.02918 &    1.14971 &    0.68588 &    0.08787 &    1.15154 &    0.14756 &    0.40497 \\

 {\bf 3.4} &    0.51320 &    0.23671 &    0.22485 &    0.02900 &    1.15583 &    0.67700 &    0.08732 &    1.13712 &    0.14670 &    0.41997 \\

 {\bf 3.5} &    0.51369 &    0.23794 &    0.22306 &    0.02891 &    1.16333 &    0.67179 &    0.08709 &    1.12887 &    0.14636 &    0.44484 \\

 {\bf 3.6} &    0.51585 &    0.23978 &    0.22342 &    0.02906 &    1.17355 &    0.67301 &    0.08756 &    1.13139 &    0.14719 &    0.47990 \\

 {\bf 3.7} &    0.51993 &    0.24246 &    0.22653 &    0.02954 &    1.18745 &    0.68249 &    0.08901 &    1.14765 &    0.14967 &    0.52127 \\

 {\bf 3.8} &    0.52604 &    0.24618 &    0.23236 &    0.03036 &    1.20566 &    0.70010 &    0.09148 &    1.17740 &    0.15386 &    0.56212 \\

 {\bf 3.9} &    0.53378 &    0.25101 &    0.24025 &    0.03144 &    1.22793 &    0.72382 &    0.09475 &    1.21720 &    0.15940 &    0.59558 \\

 {\bf 4.0} &    0.54210 &    0.25678 &    0.24917 &    0.03268 &    1.25256 &    0.75062 &    0.09849 &    1.26198 &    0.16570 &    0.61740 \\

 {\bf 4.1} &    0.54956 &    0.26315 &    0.25817 &    0.03395 &    1.27703 &    0.77758 &    0.10233 &    1.30687 &    0.17219 &    0.62672 \\

 {\bf 4.2} &    0.55511 &    0.26978 &    0.26652 &    0.03517 &    1.29937 &    0.80262 &    0.10601 &    1.34841 &    0.17837 &    0.62540 \\

 {\bf 4.3} &    0.55866 &    0.27654 &    0.27384 &    0.03625 &    1.31924 &    0.82453 &    0.10927 &    1.38465 &    0.18385 &    0.61639 \\

 {\bf 4.4} &    0.56091 &    0.28349 &    0.27983 &    0.03710 &    1.33776 &    0.84246 &    0.11183 &    1.41423 &    0.18816 &    0.60225 \\
\hline
\end{tabular}
\end{table*}

\subsection{Comparison to Previous Work}

We compare our effective collision strength results with those from previous calculations of
similar quality, that is those which used close-coupling techniques and computed collision
strengths at sufficient energies to delineate resonances.

In Table~\ref{LennonTable} we compare our final 72-term results with the $LS$ results of
\citet{LennonB1994}. That calculation was based on the 12-state target including $n=3$ correlation
orbitals described by \citet{BLS1989}. They agree within 10\% for all transitions and all
temperatures. The effective collision strengths, $\Upsilon(^3$P -- $^1$D) and $\Upsilon(^3$P --
$^1$S), for excitation of the optical forbidden lines do not differ by more than 6\% at any
temperature. The agreement is generally even better with our 20-term calculation which might be
expected since that calculation includes the 12 terms of the $n=2$ complex which is the target of
\citet{LennonB1994}. However, their target does not include the states constructed from correlation
orbitals that make a large contribution to the polarizability of the important states, as discussed
in section~\ref{target}.

The most recent R-matrix calculations where fine-structure collision strengths are presented are
those of \citet{AggarwalK1999} and \citet{PalayNPE2012}. The former calculation is based on an
elaborate 26-term target described by \citet{AggarwalH1991} constructed from 1s, 2s and 2p
spectroscopic and 3s, 3p, 3d, 4s, 4p and 4d correlation orbitals. The resulting $LS$ coupled
reactance matrices were recoupled algebraically using the JAJOM \citep{Saraph1978} program where
necessary. This approach neglects the fine-structure interactions between target states and in this
approximation some fine-structure collision strengths can be derived directly from $LS$-coupled
collision strengths using only statistical weight factors as described by both
\citet{AggarwalK1999} and \citet{LennonB1994}. \citet{PalayNPE2012} have made a 19-level
Breit-Pauli R-matrix calculation where the target is expanded over a configuration set involving
1s, 2s, 2p and 3p spectroscopic orbitals and 3p, 3d, 4s and 4p correlation orbitals.
\citet{PalayNPE2012} use an extended version of the Breit-Pauli R-matrix code which they attribute
to Eissner \& Chen (in preparation) which includes two-body fine-structure interactions which
enables them to calculate the fine structure splitting of the ground $^3$P$_J$ levels with an error
of order 3\% \citep{PalayNPE2012}. \citet{PalayNPE2012} were also the first to extend the
tabulation of thermally averaged collision strengths down to very low electron temperatures
(100~K).

In Figure~\ref{PercentCompareAllFig} we compare graphically our fine-structure results with those
of \citet{LennonB1994}, \citet{AggarwalK1999} and \citet{PalayNPE2012}. In
Table~\ref{TableCompareAll} we compare the same results numerically and also include the results of
the earlier R-matrix calculation by \citet{Aggarwal1983}. Figure~\ref{PercentCompareAllFig} shows
the percentage difference in the thermally averaged collision strengths from these three
calculations relative to our results, for all ten transitions among the energetically lowest five
levels. Where necessary, we derived fine-structure collision strengths from the results of
\citet{LennonB1994} and \citet{AggarwalK1999} using statistical weight factors as outlined above.
With the exception of the \SLPJ1De2 -- \SLPJ1Se0 transition, our results agree with those of
\citet{LennonB1994} and \citet{AggarwalK1999} to within 10\% for all temperatures between 1000~K
and 25000~K where comparison can be made and to within 5\% for the majority of temperatures. For
these two calculations the differences are relatively insensitive to temperature, indicating that
their collision strengths have a similar energy dependence to ours. We find generally larger
disagreements with the results of \citet{PalayNPE2012}, reaching 10--15\% at the extremes of
tabulated temperature for many transitions and being even larger for the transitions from the
ground $^3$P$_J$ levels to the $^1$S$_0$ state (transitions 1-5, 2-5 and 3-5). Here the differences
reach 100\% at 100~K and are over 20\% at 10000~K. The differences also show a distinctive
temperature dependence. With the exception of the $^3$P$_J$ -- $^1$S$_0$ transitions the
\citet{PalayNPE2012} results are generally smaller than ours at the lowest temperatures and larger
at the highest temperatures. This suggests that their collision strengths generally have a
different energy dependence in the energy range relevant for nebular temperatures, about
1/4~Rydberg above threshold.

\subsection{Discussion}

\begin{figure}
\centering{}
\includegraphics[scale=0.65]{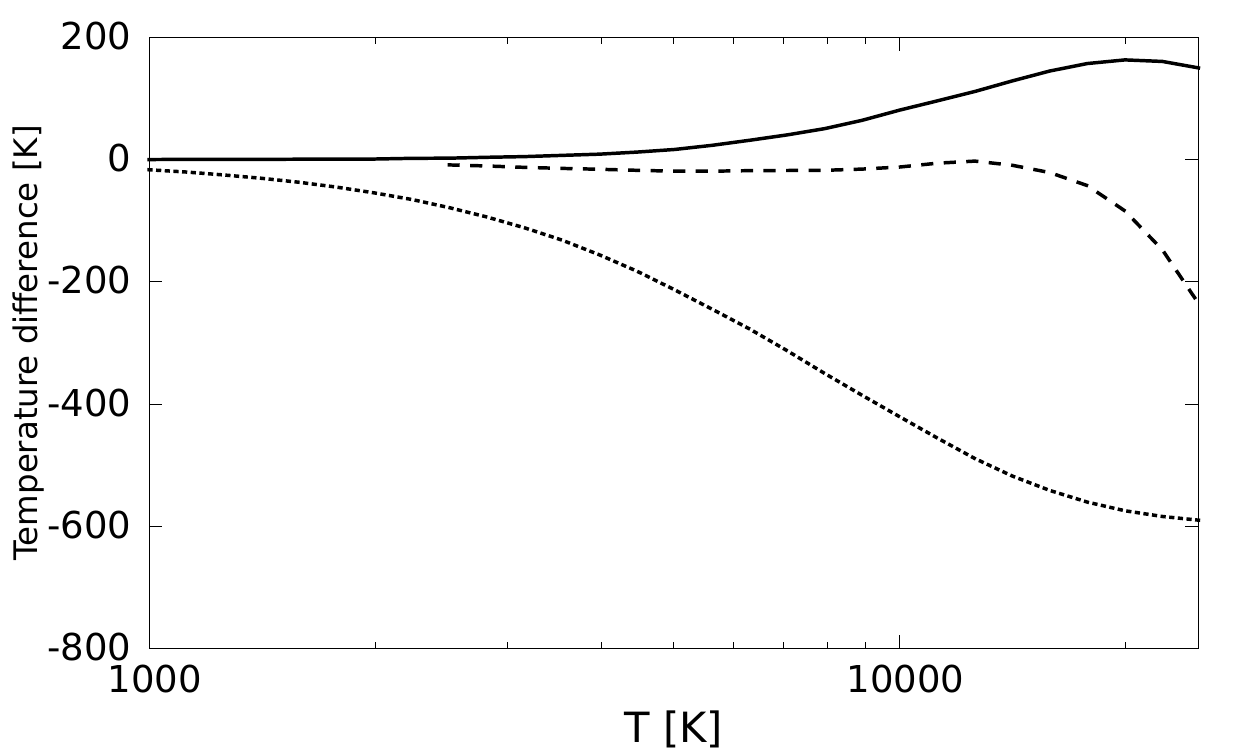}
\caption{The difference in derived electron temperature from the
$\lambda$4363/($\lambda$4959+$\lambda$5007) line intensity ratio using the data of
 \citet{LennonB1994} (solid line), \citet{AggarwalK1999} (dashed line) and  \citet{PalayNPE2012}
(dotted line) against the temperature derived from the present results. } \label{tvsdt}
\end{figure}

In photoionized plasmas the O~{\sc iii} forbidden lines are commonly used to determine the
electron temperature of the emitting material, and hence to determine the number of O$^{2+}$
emitters relative to H by comparison with a strong H recombination line. The temperature
determination rests on the ratio of the intensity of the $\lambda4363$ line to either or both of
the $\lambda4959$ and $\lambda5007$ lines. The $\lambda4363$ line is relatively weak and cannot be
seen if the temperature is much below 5000~K. Once the temperature is known, the much stronger
$\lambda\lambda4959, 5007$ lines can be used to deduce the O$^{2+}$ number density. In nebular
plasmas all these lines are excited collisionally from the $^3$P$_J$ ground levels. The excitation
mechanism for $\lambda4363$ is therefore central to determining the electron temperature and
abundances. In Figure~\ref{tvsdt} we show how the derived electron temperature from our work
differs from that obtained from \citet{LennonB1994} and from the data of \citet{AggarwalK1999} and
\citet{PalayNPE2012}. In all the temperature determinations the radiative transition probabilities
were taken from \citet{NussbaumerS1981} and \citet{StoreyZ2000}.  Very similar temperatures are
obtained with the collision strength data of \citet{LennonB1994} and \citet{AggarwalK1999}.
\citet{PalayNPE2012} state that there are no significant differences in line ratios arising from
their calculation when comparing to \citet{AggarwalK1999} but Figure~\ref{tvsdt} shows that this is
not the case. The difference in derived temperature is 213~K at 5000~K, 421~K at 10000~K and 504~K
at 15000~K.

\begin{figure}
\centering{}
\includegraphics[scale=0.65]{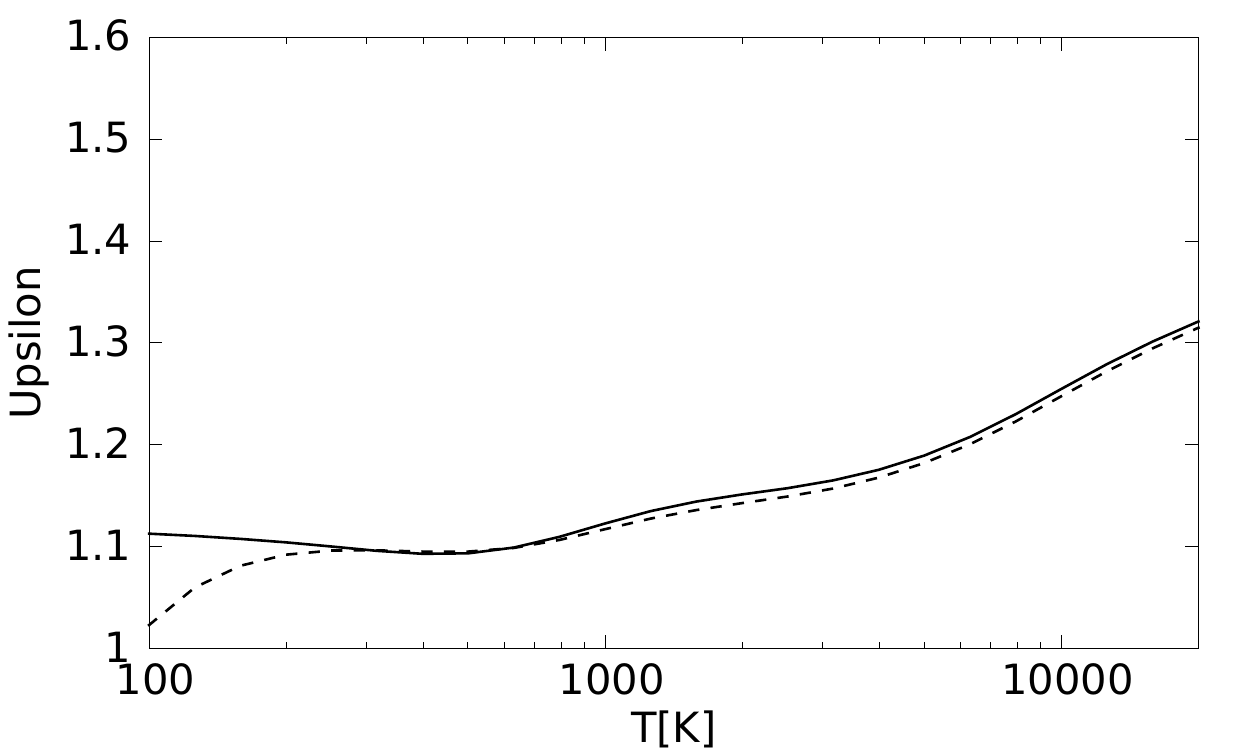}
\caption{Effective collision strength (Upsilon) versus temperature from 72-term target: ICFT with
spin-orbit interactions in the target (solid line) and ICFT in pair coupling (dashed line) for the
transition between levels 2 and 3 of Table \ref{levelsTable}.} \label{compicjk}
\end{figure}

\begin{figure}
\centering{}
\includegraphics[scale=0.45]{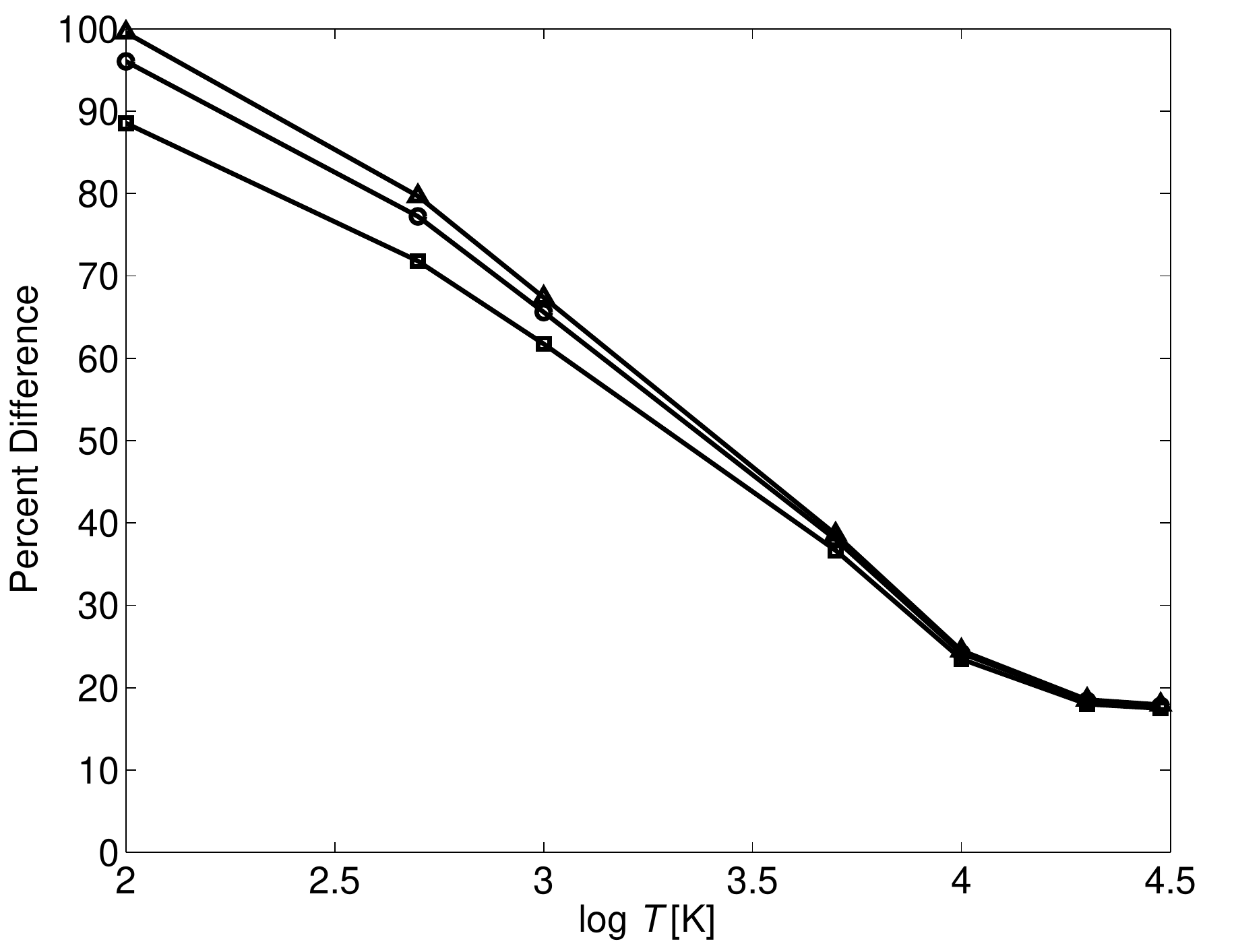}
\caption{Percent difference in $\Upsilon$ versus logarithm of temperature between the results of \citet{PalayNPE2012} and our 72-term Breit-Pauli
calculation for the transitions
\SLPJ3Pe0-\SLPJ1Se0 (triangle), \SLPJ3Pe1-\SLPJ1Se0 (circle) and \SLPJ3Pe2-\SLPJ1Se0 (square).
\label{PalayFigure2All}}
\end{figure}

In summary, our new Breit-Pauli R-matrix calculation generally shows
much better agreement for thermally averaged collision strength with
the earlier non-Breit-Pauli R-matrix calculations of
\citet{LennonB1994} and \citet{AggarwalK1999} than the more recent
Breit-Pauli work of \citet{PalayNPE2012}. The results of the important
forbidden line diagnostic line ratios show the same pattern. The
reasons why the \citet{PalayNPE2012} results differ is not clear. One
question that arises is whether the two-body fine-structure terms that
are included in the Breit-Pauli R-matrix formulation of
\citet{PalayNPE2012} and not in our calculation might be the cause. We
do not believe that this is the case for the following reason. In
Figure~\ref{compicjk} we show two sets of results for the thermally averaged
collision strength for the \SLPJ3Pe1 -- \SLPJ3Pe2 transition from
72-term ICFT calculations. The solid line includes the effects of the
spin-orbit interaction in the target, introduced {\it via} the
so-called Term-Coupling Coefficients (TCCs), while the dashed line
shows the results obtained in pair-coupling, i.e. without TCCs. Except
at the lowest temperatures ($T < 300$~K) they differ by no more than
1\%.  The larger difference at the lowest temperatures simply reflects
the fact that the pair-coupling calculation does not separate the
$^3$P$_J$ levels in energy and therefore the threshold energies of
these levels are not correct. The results for the other transitions
show similar behavior. We emphasize, however, that the ICFT
calculation which does incorporate target spin-orbit effects agrees
with the full Breit-Pauli calculation to within 1\% at all
temperatures.  The good agreement that we find shows that the
spin-orbit interaction has a very small effect on the results. In
O$^{2+}$ two-body fine-structure interactions are substantially
smaller than the spin-orbit interaction and should therefore have a
negligible effect on the results. This point is emphasized in
Figure~\ref{PalayFigure2All} where we show the percentage difference
between the results of \citet{PalayNPE2012} and ours for the three
$^3$P$_J$ -- $^1$S$_0$ transitions. Except at very low temperatures,
they do not show any significant dependence on $J$ which might be
expected if fine-structure effects were important and indicate rather
that the term-term $^3$P -- $^1$S collision strengths differ
significantly between the two calculations.

\begin{figure*}
\centering %
\subfigure[1-2]%
{
\begin{minipage}[b]{0.32\textwidth}
\CIF {g/PCA12_2}
\end{minipage}}
\Hs %
\subfigure[1-3]%
{
\begin{minipage}[b]{0.32\textwidth}
\CIF {g/PCA13_2}
\end{minipage}}
\Hs
\subfigure[1-4]%
{
\begin{minipage}[b]{0.32\textwidth}
\CIF {g/PCA14_2}
\end{minipage}} \Vmin
%
\centering %
\subfigure[1-5]%
{
\begin{minipage}[b]{0.32\textwidth}
\CIF {g/PCA15_2}
\end{minipage}}
\Hs
\subfigure[2-3]%
{
\begin{minipage}[b]{0.32\textwidth}
\CIF {g/PCA23_2}
\end{minipage}}
\Hs
\subfigure[2-4]%
{\begin{minipage}[b]{0.32\textwidth} \CIF {g/PCA24_2}
\end{minipage}} \Vmin
%
\centering %
\subfigure[2-5]%
{\begin{minipage}[b]{0.32\textwidth} \CIF {g/PCA25_2}
\end{minipage}} \Vmin
\Hs
\subfigure[3-4]%
{\begin{minipage}[b]{0.32\textwidth} \CIF {g/PCA34_2}
\end{minipage}}
\Hs
\subfigure[3-5]%
{\begin{minipage}[b]{0.32\textwidth} \CIF {g/PCA35_2}
\end{minipage}} \Vmin
%
\centering %
\subfigure[4-5]%
{\begin{minipage}[b]{0.32\textwidth} \CIF {g/PCA45_2}
\end{minipage}}
\caption{Percentage differences of thermally averaged collision
  strengths from our 72-term Breit-Pauli
  calculation, (vertical axis) versus temperature in 10000~K
  (horizontal axis). Results from \citet{LennonB1994} (solid line),
  \citet{AggarwalK1999} (dashed line) and \citet{PalayNPE2012}
  (dotted line). The labels of the sub-figures refer to the level
  indices in Table \ref{levelsTable}. \label{PercentCompareAllFig}}
\end{figure*}

\section{Conclusions} \label{Conclusions}

In the present paper, the collision strengths for the transitions between the lowest five levels of
the astronomically-important ${\rm O}^{2+}+e^{-}$ atomic system up to about 1.3~Rydberg of electron
excitation energy are computed in the close coupling approximation using the
UCL-Belfast-Strathclyde R-matrix atomic code. Different coupling schemes with different atomic
definitions and parameters are used to describe the scattering target and scattering process.

Our results were extensively compared to previous work. We found a good agreement in most cases
which increases our confidence in our results. However, we found significant differences with
\citet{PalayNPE2012} who also used a Breit-Pauli coupling scheme and hence a better agreement was
expected. The good agreement between our R-matrix Breit-Pauli calculation and earlier R-matrix work
in which the fine-structure was treated more approximately strongly supports our results. We showed
that the relatively large differences found for the excitation of the $\lambda4363$ line between
the work of \citet{PalayNPE2012} on the one hand, and all previous calculations, on the other,
leads to significant differences in derived temperatures from the main [O~{\sc iii}] line ratios.

With regard to the use of the ICFT method, for lowly ionized systems some resonances can have very
low principal quantum number, and channels are deeply closed, which can cause problems for
multichannel quantum defect theory. This difficulty can be overcome by explicitly omitting channels
with very low effective quantum number and in any case evaporates as the effective charge number
increases.

\section{Acknowledgments and Statement}

The work of PJS and NRB was supported in part by STFC (grant ST/J000892/1). Full-precision data for
the energy-dependent collision strengths of the transitions between the lowest five levels of the
investigated ${\rm O}^{2+}+e^{-}$ system using the 72-term target under a Breit-Pauli intermediate
coupling scheme can be obtained in electronic format from the Centre de Donn\'{e}es astronomiques
de Strasbourg database.

\onecolumn

\begin{table}
\caption{Comparison of thermally averaged collision strengths, $\Upsilon$, from \citet{LennonB1994} (LB) and the current work within the lowest five levels as a function of temperature. The first row of each
temperature is from Table~3 of LB and the second row is from the current work using the Breit-Pauli
72-term target.\label{LennonTable}}
\begin{center}
\begin{tabular}{lcccccc}
\hline
    log~$T$~[K] &    \SLPJ3Pe0-\SLPJ3Pe1 &    \SLPJ3Pe0-\SLPJ3Pe2 &    \SLPJ3Pe1-\SLPJ3Pe2 &     \SLP3Pe-\SLPJ1De2 &     \SLP3Pe-\SLPJ1Se0 &    \SLPJ1De2-\SLPJ1Se0 \\
\hline
       3.0 &     0.4975 &     0.2455 &     1.1730 &     2.2233 &     0.2754 &     0.4241 \\

           &     0.5199 &     0.2313 &     1.1218 &     2.1331 &     0.2667 &     0.3897 \\
\hline
       3.2 &     0.5066 &     0.2493 &     1.1930 &     2.1888 &     0.2738 &     0.4268 \\

           &     0.5154 &     0.2349 &     1.1430 &     2.0811 &     0.2643 &     0.3968 \\
\hline
       3.4 &     0.5115 &     0.2509 &     1.2030 &     2.1416 &     0.2713 &     0.4357 \\

           &     0.5132 &     0.2367 &     1.1558 &     2.0237 &     0.2610 &     0.4200 \\
\hline
       3.6 &     0.5180 &     0.2541 &     1.2180 &     2.1117 &     0.2693 &     0.4652 \\

           &     0.5158 &     0.2398 &     1.1736 &     2.0107 &     0.2616 &     0.4799 \\
\hline
       3.8 &     0.5296 &     0.2609 &     1.2480 &     2.1578 &     0.2747 &     0.5232 \\

           &     0.5260 &     0.2462 &     1.2057 &     2.0913 &     0.2732 &     0.5621 \\
\hline
       4.0 &     0.5454 &     0.2713 &     1.2910 &     2.2892 &     0.2925 &     0.5815 \\

           &     0.5421 &     0.2568 &     1.2526 &     2.2425 &     0.2941 &     0.6174 \\
\hline
       4.2 &     0.5590 &     0.2832 &     1.3350 &     2.4497 &     0.3174 &     0.6100 \\

           &     0.5551 &     0.2698 &     1.2994 &     2.3987 &     0.3165 &     0.6254 \\
\hline
       4.4 &     0.5678 &     0.2955 &     1.3730 &     2.5851 &     0.3405 &     0.6090 \\

           &     0.5609 &     0.2835 &     1.3378 &     2.5184 &     0.3339 &     0.6022 \\
\hline

\end{tabular}
\end{center}
\end{table}

\begin{table}
  \caption{Comparison of thermally averaged collision strengths, $\Upsilon$, between \citet{Aggarwal1983} (A),
    \citet{LennonB1994} (LB), \citet{AggarwalK1999} (AK), \citet{PalayNPE2012} (P), and the current
    work (SSB) using the 72-term target as a function of temperature [K]. The transitions are  indexed
    as in Table \ref{levelsTable}. Note that the values attributed to \citet{LennonB1994} for $T=2500$K and $25000$K are those tabulated for log~$T = 3.4$, $4.4$ in that work. \label{TableCompareAll}}
\begin{center}
\begin{tabular}{|l@{  }|l@{  }|c@{  }|c@{  }|c@{  }|c@{  }|c@{  }|c@{  }|c@{  }|c@{  }|c@{  }|c@{  }|c@{  }|c@{  }|c|}
\hline
      Index &           &                                                                                                                                  \multicolumn{ 13}{|c}{Temperature [K]} \\
\hline
           &            &        100 &        500 &       1000 &       2500 &       5000 &       7500 &      10000 &      12500 &      15000 &      17500 &      20000 &      25000 &      30000 \\
\hline
        1-2 &          A &            &            &            &     0.5041 &     0.5172 &     0.5310 &     0.5417 &     0.5490 &     0.5537 &     0.5567 &     0.5586 &     0.5612 &     0.5633 \\

            &        LB &            &            &     0.4975 &     0.5115 &            &            &     0.5454 &              &            &            &            &      0.5678 &            \\

           &         AK &            &            &            &     0.5011 &     0.5084 &     0.5159 &     0.5222 &     0.5266 &     0.5294 &     0.5311 &     0.5324 &     0.5348 &     0.5380 \\

           &          P &     0.5814 &     0.5005 &     0.4866 &            &     0.5240 &            &     0.5648 &            &            &            &     0.6007 &            &     0.6116 \\

           &        SSB &     0.6350 &     0.5430 &     0.5199 &     0.5132 &     0.5199 &     0.5317 &     0.5421 &     0.5494 &     0.5540 &     0.5569 &     0.5587 &     0.5609 &     0.5623 \\
\hline
        1-3 &          A &            &            &            &     0.2499 &     0.2566 &     0.2646 &     0.2717 &     0.2776 &     0.2824 &     0.2865 &     0.2901 &     0.2962 &     0.3013 \\

           &         LB &            &            &      0.2455 &     0.2509 &            &            &     0.2713 &            &            &            &            &      0.2955 &            \\

           &         AK &            &            &            &     0.2406 &     0.2449 &     0.2512 &     0.2573 &     0.2626 &     0.2669 &     0.2707 &     0.2739 &     0.2798 &     0.2855 \\

           &          P &     0.2142 &     0.2153 &     0.2234 &            &     0.2469 &            &     0.2766 &            &            &            &     0.3106 &            &     0.3264 \\

           &        SSB &     0.2259 &     0.2247 &     0.2313 &     0.2367 &     0.2424 &     0.2497 &     0.2568 &     0.2629 &     0.2682 &     0.2727 &     0.2766 &     0.2833 &     0.2890 \\
\hline
        1-4 &          A &            &            &            &     0.2283 &     0.2262 &     0.2337 &     0.2426 &     0.2506 &     0.2627 &     0.2627 &     0.2672 &     0.2740 &     0.2790 \\

            &        LB &            &            &      0.2470 &     0.2380 &            &            &     0.2544 &             &           &            &            &      0.2872 &            \\

           &         AK &            &            &            &     0.2260 &     0.2265 &     0.2343 &     0.2434 &     0.2515 &     0.2582 &     0.2637 &     0.2683 &     0.2751 &     0.2799 \\

           &          P &     0.1959 &     0.2088 &     0.2154 &            &     0.2347 &            &     0.2693 &            &            &            &     0.3094 &            &     0.3256 \\

           &        SSB &     0.2318 &     0.2389 &     0.2370 &     0.2249 &     0.2265 &     0.2381 &     0.2492 &     0.2579 &     0.2646 &     0.2698 &     0.2739 &     0.2797 &     0.2832 \\
\hline
        1-5 &          A &            &            &            &     0.0278 &     0.0280 &     0.0295 &     0.0310 &     0.0324 &     0.0335 &     0.0344 &     0.0351 &     0.0362 &     0.0368 \\

            &        LB &            &            &      0.0306 &     0.0301 &            &            &     0.0325 &             &           &            &            &      0.0378 &            \\

           &         AK &            &            &            &     0.0307 &     0.0304 &     0.0310 &     0.0321 &     0.0332 &     0.0342 &     0.0351 &     0.0358 &     0.0370 &     0.0378 \\

           &          P &     0.0597 &     0.0535 &     0.0496 &            &     0.0409 &            &     0.0407 &            &            &            &     0.0430 &            &     0.0442 \\

           &        SSB &     0.0299 &     0.0298 &     0.0296 &     0.0290 &     0.0295 &     0.0312 &     0.0327 &     0.0339 &     0.0349 &     0.0357 &     0.0363 &     0.0371 &     0.0375 \\
\hline
        2-3 &          A &            &            &            &     1.1925 &     1.2239 &     1.2592 &     1.2884 &     1.3107 &     1.3275 &     1.3404 &     1.3510 &     1.3679 &     1.3821 \\

            &        LB &            &            &     1.1730 &     1.2030 &            &            &     1.2910  &             &           &            &            &     1.3730 &            \\

           &         AK &            &            &            &     1.1680 &     1.1870 &     1.2100 &     1.2320 &     1.2490 &     1.2620 &     1.2730 &     1.2820 &     1.2980 &     1.3150 \\

           &          P &     1.0360 &     1.0320 &     1.0720 &            &     1.2100 &            &     1.3300 &            &            &            &     1.4510 &            &     1.4990 \\

           &        SSB &     1.1121 &     1.0928 &     1.1218 &     1.1557 &     1.1873 &     1.2221 &     1.2526 &     1.2763 &     1.2943 &     1.3082 &     1.3194 &     1.3374 &     1.3518 \\
\hline
        2-4 &          A &            &            &            &     0.6848 &     0.6785 &     0.7010 &     0.7279 &     0.7518 &     0.7716 &     0.7879 &     0.8014 &     0.8221 &     0.8368 \\

            &        LB &            &            &      0.7411 &     0.7139 &            &            &     0.7631 &             &           &            &            &      0.8617 &            \\

           &         AK &            &            &            &     0.6780 &     0.6795 &     0.7029 &     0.7302 &     0.7545 &     0.7746 &     0.7911 &     0.8049 &     0.8253 &     0.8397 \\

           &          P &     0.5903 &     0.6285 &     0.6483 &            &     0.7067 &            &     0.8108 &            &            &            &     0.9313 &            &     0.9802 \\

           &        SSB &     0.6975 &     0.7187 &     0.7132 &     0.6772 &     0.6823 &     0.7175 &     0.7506 &     0.7768 &     0.7969 &     0.8125 &     0.8247 &     0.8421 &     0.8527 \\
\hline
        2-5 &          A &            &            &            &     0.0833 &     0.0840 &     0.0884 &     0.0931 &     0.0972 &     0.1006 &     0.1033 &     0.1054 &     0.1085 &     0.1105 \\

            &        LB &            &            &     0.0918 &     0.0904  &            &            &     0.0975 &             &           &            &            &     0.1135 &             \\

           &         AK &            &            &            &     0.0921 &     0.0911 &     0.0929 &     0.0962 &     0.0995 &     0.1025 &     0.1052 &     0.1074 &     0.1109 &     0.1135 \\

           &          P &     0.1765 &     0.1590 &     0.1477 &            &     0.1228 &            &     0.1223 &            &            &            &     0.1294 &            &     0.1332 \\

           &        SSB &     0.0900 &     0.0897 &     0.0892 &     0.0873 &     0.0890 &     0.0939 &     0.0985 &     0.1022 &     0.1052 &     0.1075 &     0.1093 &     0.1118 &     0.1131 \\
\hline
        3-4 &          A &            &            &            &     1.1413 &     1.1308 &     1.1683 &     1.2131 &     1.2529 &     1.2860 &     1.3132 &     1.3357 &     1.3702 &     1.3947 \\

           &         LB &            &            &     1.2352 &     1.1898 &            &            &     1.2718  &            &            &            &            &     1.4362 &            \\

           &         AK &            &            &            &     1.1300 &     1.1325 &     1.1715 &     1.2170 &     1.2575 &     1.2910 &     1.3185 &     1.3415 &     1.3755 &     1.3995 \\

           &          P &     0.9934 &     1.0560 &     1.0890 &            &     1.1880 &            &     1.3630 &            &            &            &     1.5640 &            &     1.6450 \\

           &        SSB &     1.1702 &     1.2057 &     1.1965 &     1.1374 &     1.1474 &     1.2066 &     1.2620 &     1.3055 &     1.3389 &     1.3647 &     1.3850 &     1.4137 &     1.4310 \\
\hline
        3-5 &          A &            &            &            &     0.1388 &     0.1401 &     0.1473 &     0.1552 &     0.1620 &     0.1676 &     0.1721 &     0.1757 &     0.1809 &     0.1842 \\

           &         LB &            &            &      0.1530 &     0.1507 &            &            &     0.1625 &            &            &            &            &      0.1892 &            \\

           &         AK &            &            &            &     0.1536 &     0.1518 &     0.1549 &     0.1603 &     0.1659 &     0.1709 &     0.1753 &     0.1790 &     0.1849 &     0.1891 \\

           &          P &     0.2850 &     0.2587 &     0.2421 &            &     0.2045 &            &     0.2046 &            &            &            &     0.2170 &            &     0.2235 \\

           &        SSB &     0.1512 &     0.1506 &     0.1497 &     0.1467 &     0.1496 &     0.1579 &     0.1657 &     0.1720 &     0.1769 &     0.1808 &     0.1839 &     0.1881 &     0.1902 \\
\hline
        4-5 &          A &            &            &            &     0.4708 &     0.5463 &     0.6114 &     0.6468 &     0.6630 &     0.6687 &     0.6692 &     0.6670 &     0.6599 &     0.6524 \\

            &        LB &            &            &     0.4241 &     0.4357 &            &            &     0.5815  &             &           &            &            &     0.6090 &            \\

           &         AK &            &            &            &     0.3907 &     0.4312 &     0.4836 &     0.5227 &     0.5478 &     0.5629 &     0.5719 &     0.5769 &     0.5809 &     0.5812 \\

           &          P &     0.3900 &     0.3899 &     0.3899 &            &     0.4544 &            &     0.5661 &            &            &            &     0.6230 &            &     0.6219 \\

           &        SSB &     0.3827 &     0.3856 &     0.3897 &     0.4196 &     0.5208 &     0.5882 &     0.6174 &     0.6266 &     0.6265 &     0.6223 &     0.6163 &     0.6026 &     0.5886 \\
\hline
\end{tabular}
\end{center}
\end{table}

\end{document}